\documentclass[conference]{IEEEtran}
\usepackage[affil-it]{authblk}

\usepackage{graphicx}
\usepackage{caption}
\usepackage{multirow}
\usepackage{textcomp}
\usepackage{adjustbox,lipsum}
\usepackage{xspace}
\usepackage{amsmath}
\usepackage{amssymb}
\usepackage{amsfonts}
\usepackage{amstext}
\usepackage{amsthm}
\usepackage{tabularx}
\usepackage{caption}
\usepackage{subcaption}
\usepackage{xcolor}
\usepackage{colortbl}
\usepackage{algorithm}
\usepackage[noend]{algpseudocode}
\usepackage{cite}
\usepackage{comment}
\usepackage{subfiles} 
\newcommand{\MethodName}{AnoMili\xspace}

\algnewcommand\algorithmicforeach{\textbf{for each}}
\algdef{S}[FOR]{ForEach}[1]{\algorithmicforeach\ #1\ \algorithmicdo}

\algdef{SE}[DOWHILE]{Do}{doWhile}{\algorithmicdo}[1]{\algorithmicwhile\ #1}%

\algdef{SE}[VARIABLES]{Variables}{EndVariables}
   {\algorithmicvariables}
   {\algorithmicend\ \algorithmicvariables}
\algnewcommand{\algorithmicvariables}{\textbf{global variables}}

\begin{document}

\title{ \MethodName : Spoofing Prevention and Explainable Anomaly Detection for the 1553 Military Avionic Bus}

\author{Efrat Levy}
\author{Nadav Maman}
\author{Asaf Shabtai}
\author{Yuval Elovici}
\affil{Department of Software and Information Systems Engineering, Ben-Gurion University of the Negev}

\date{Dated: \today}

\maketitle

\begin{abstract}
MIL-STD-1553, a standard that defines a communication bus for interconnected devices, is widely used in military and aerospace avionic platforms.
Due to its lack of security mechanisms, MIL-STD-1553 is exposed to cyber threats. The methods previously proposed to address these threats are very limited, resulting in the need for more advanced techniques.
Inspired by the \emph{defense in depth} principle, we propose \MethodName, a novel protection system for the MIL-STD-1553 bus, which consists of: (i) a physical intrusion detection mechanism that detects unauthorized devices connected to the 1553 bus, even if they are passive (sniffing), (ii) a device fingerprinting mechanism that protects against spoofing attacks (two approaches are proposed: prevention and detection), (iii) a context-based anomaly detection mechanism, and (iv) an anomaly explanation engine responsible for explaining the detected anomalies in real time.
We evaluate \MethodName's effectiveness and practicability in two real 1553 hardware-based testbeds. The effectiveness of the anomaly explanation engine is also demonstrated.
All of the detection and prevention mechanisms employed had high detection rates (over 99.45\%) with low false positive rates. 
The context-based anomaly detection mechanism obtained perfect results when evaluated on a dataset used in prior work. 
\end{abstract}
\vspace{0.5cm}
\section{\label{sec:intro}Introduction}

MIL-STD-1553 is a military standard that defines a real-time communication bus for interconnected devices.
Published by the US Department of Defense (DoD) in 1973, it is widely used in military and aerospace avionic platforms (e.g., F-35 and F-16)~\cite{bracknell2007mil}. 
MIL-STD-1553 defines both the physical and logical requirements for implementing the 1553 bus and focuses on providing a high level of fault tolerance~\cite{mil19981553}. 
Despite its importance, the 1553 bus was designed without security features, making the entire 1553 system susceptible to modern cyber threats that can compromise the confidentiality, integrity, and availability of systems that use the 1553 bus~\cite{stan2018security,de2021exploiting,nguyen2015towards}. 

\begin{figure}[h]
\scriptsize
\centering
\includegraphics[width=0.475\textwidth]{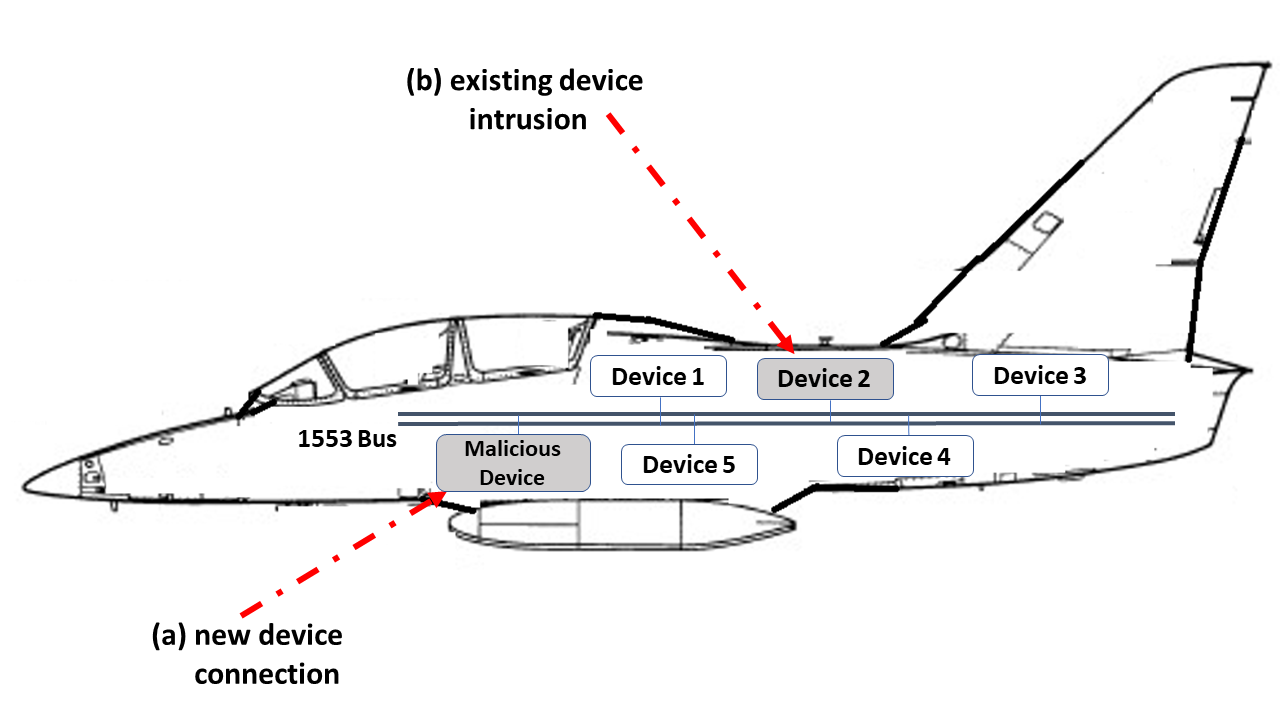}
\caption{Attack surfaces of a platform using the MIL-STD-1553 bus.}
\label{fig:attack_surfaces}
\end{figure}

Common security mechanisms (e.g., firewalls, malware detection, data leakage prevention, and access control) are not suitable for the 1553 bus. 
Besides the computational overhead, they require significant adaptation, and many legacy systems that cannot be changed are connected to the 1553 bus.

Several studies examined ways of securing the 1553 that do not necessitate changes to the operating systems or communication protocol; in those studies, statistical methods for detecting anomalies in the transmitted messages were proposed ~\cite{losier2019design,genereux2019maidens,stan2019intrusion,onodueze2020anomaly}.
However, none of the studies focused on prevention or providing an explanation for the anomalies detected. 
The ability to efficiently and automatically explain the anomalies is necessary for real-time response and remediation. 

In this paper, we present \MethodName, an end-to-end security system for the 1553 bus that provides real-time anomaly explainability. 
Inspired by the \emph{defense in depth} principle~\cite{kuipers2006control,jajodia2011cauldron}, \MethodName consists of two phases. In the first phase, \MethodName hardens the bus against insider threats and utilizes physical side-channels to immediately notify the operational staff if a new (potentially malicious) device is connected to the bus (i.e., the bus is physically compromised). 
This is done using an unsupervised deep learning-based mechanism which analyzes the legitimate devices' voltage signals measured on the bus; this mechanism, which we refer to as the \textit{physical intrusion detection} mechanism, is executed when the aircraft starts; this mechanism is also effective against silent devices.
If no new connected devices are detected, in the next phase, \MethodName continuously monitors the messages transmitted on the bus and hardens the bus using the following mechanisms: the \textit{device fingerprinting} and \textit{context-based anomaly detection} mechanisms.

The goal of the device fingerprinting mechanism is to address spoofing scenarios. In this paper, we propose two spoofing protection approaches: detection and prevention.
The \textit{detection} approach uses deep learning-based classifiers to analyze the unique characteristics of the voltage signals measured on the bus during a message transmission and authenticate the origin device. Since voltage signals fluctuate over time due to environmental changes, these classifiers are continuously updated.
The \textit{prevention} approach is implemented as a wrapper for the basic 1553 hardware transceiver.
This wrapper is responsible for efficiently preventing spoofing attempts originating from any software component running on a device; it does this by comparing the source address in a message during a message writing attempt with the real (known) source address of the device.
While this solution requires changes to the hardware of each transceiver, this solution is seamless to the system running above and adds negligible computational overhead.

The context-based anomaly detection mechanism is aimed at identifying anomalous messages based on the transmission context. 
This is done by using an unsupervised deep learning algorithm to model sequences of messages and identify anomalous messages. 

In order to assist \MethodName's users in understanding the alerts and taking the correct action, we propose an anomaly explanation engine. This engine is responsible for explaining the detected anomalies in real time. Each explanation is represented at a high level of abstraction; this is used by the pilot, and it contains information on the attack vector (e.g., device \emph{i} is compromised) and a description of the attack (e.g., transmission of a firing command message followed by a fake location message).
When an anomalous message is detected by the context-based anomaly detection mechanism, the anomaly explanation engine also provides an anomaly explanation at a low level of abstraction; it calculates the features of the anomalous message that most influence the mechanism's prediction (e.g., message length or source address).
To the best of our knowledge, this is the first study in the transportation domain to design a real-time mechanism that produces human-actionable insights regarding the anomalies detected.

To evaluate \MethodName, we created two testbeds based on real 1553 hardware, within which we implemented 10 physical and logical attack scenarios. 
The physical intrusion detection mechanism demonstrated perfect detection accuracy (i.e., in each experiment performed, the new devices connected to the bus were detected) with zero false positives.
The detection approach of the device fingerprinting mechanism obtained over 99.45\% classification accuracy, and the prevention approach was able to block unauthorized bus writing in all scenarios examined.

The context-based anomaly detection mechanism demonstrated perfect results (all anomalous messages were detected with zero false alarms) for both normal and abnormal scenarios when evaluated using datasets collected from our two testbeds and the dataset used by Stan et al.~\cite{stan2019intrusion}. 
In addition, we demonstrated the ability of the anomaly explanation engine to accurately explain the anomalies. 

Besides voltage signals-based detection mechanisms, we show that all the \MethodName's mechanisms are transferable from one 1553 system to another 1553 system without retraining. 
Regarding the voltage signals-based detection mechanisms, we report that a few minutes of training are sufficient for generating the machine learning models.

To summarize, the main contributions of this paper are as follows:
\begin{itemize}
    \item A mechanism for detecting unauthorized devices connected to the 1553 bus (i.e., physical intrusions), which is effective even when the connected devices are silent.
    \item A mechanism for detecting spoofing attempts that can adapt to environmental changes.
    \item A mechanism for preventing spoofing attempts that does not require any changes to the operating system or communication protocol.
    \item A mechanism for detecting anomalous messages based on their context during data transmission, whose predictions are feasible to explain. 
    \item A real-time anomaly explanation engine that automatically generates practical/actionable explanations for the anomalies detected. 
    \item An evaluation conducted on two real 1553 hardware-based testbeds, as well as on a dataset that was used in prior work~\cite{stan2019intrusion}. 
    \item Most of the proposed mechanisms in this study are transferable from one 1553 system to another 1553 system without retraining. The rest only require a few minutes of models training.
\end{itemize}
\vspace{0.5cm}

\section{Background: Military Avionics Systems}
\subsection{Military Avionics Functions}
Military avionics is a tactical version of avionics, focused on electronic systems and equipment used in military aircraft of all kinds. These include flight control and navigation functions similar to those in commercial aircraft, as well as electro-optic and infrared threat sensors, activity monitors, secure tactical communications, weapons trackers, countermeasure capabilities, and other integrated electronic support and protection capabilities. 
Those systems all communicate through the 1553 bus, and most of them include both a status/information reporter unit and an internal entity that expects to receive operational commands.

\begin{figure}[h]
\scriptsize
\centering
\includegraphics[width=0.50\textwidth]{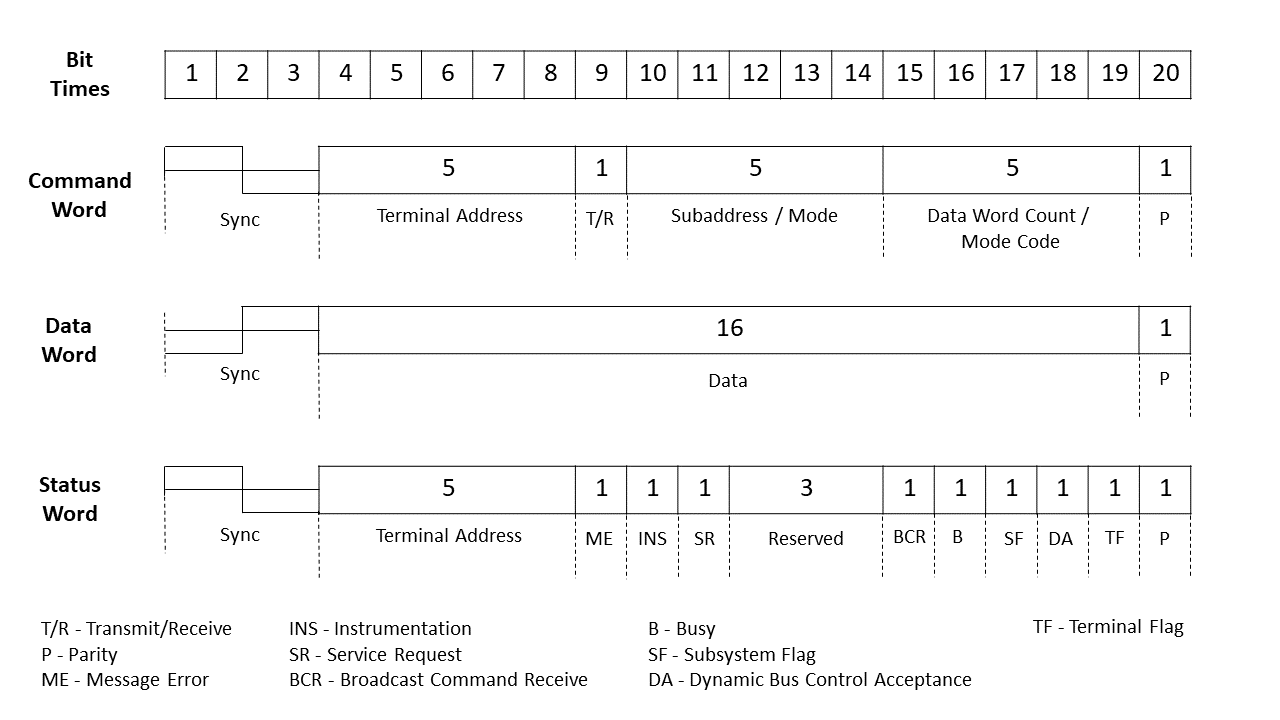}
\caption{MIL-STD-1553 word formats.}
\label{fig:word_formats}
\end{figure}

\begin{figure}[h]
\scriptsize
\centering
\includegraphics[width=0.47\textwidth]{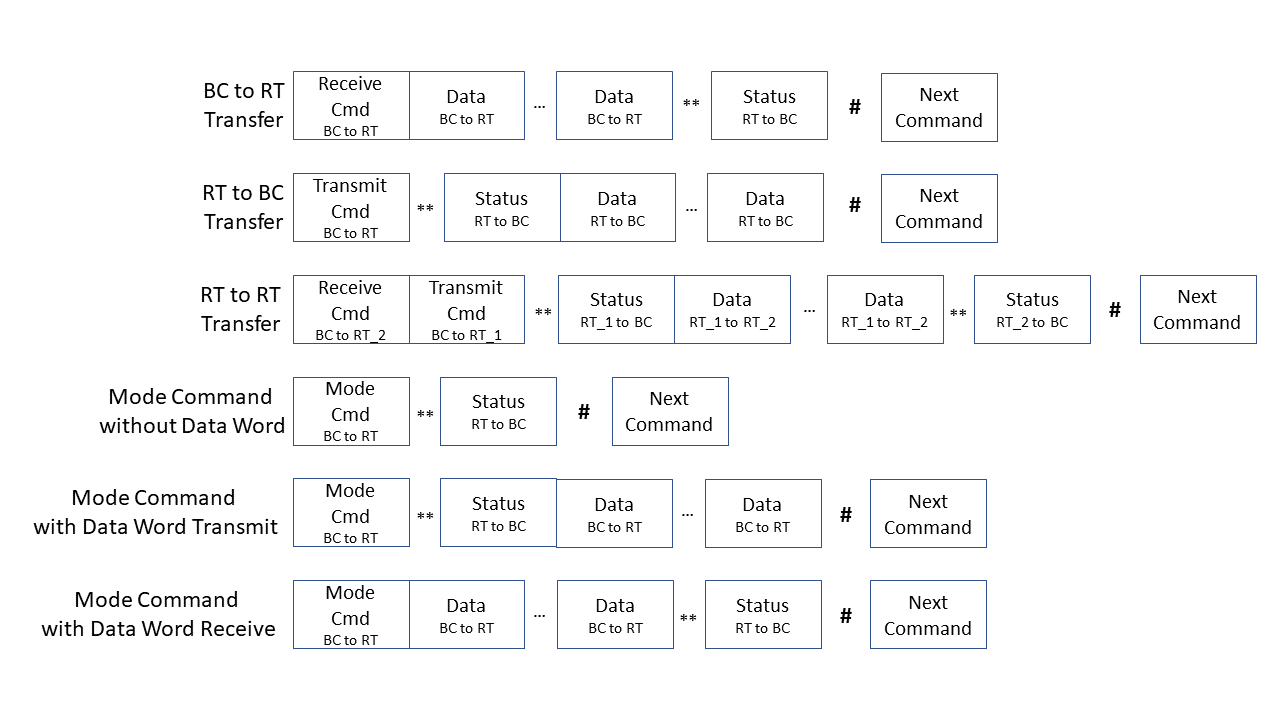}
\caption{MIL-STD-1553 message transfer formats.}
\label{fig:information_transfer_formats}
\end{figure}

\vspace{0.5cm}
\subsection{The MIL-STD-1553 Communication Bus}
MIL-STD-1553 defines the requirements for digital, command/response, and time division multiplexing techniques for a dual redundant 1-MHz serial bus and specifies the communication bus and its electronic interface. 
All transmissions on the 1553 bus are accessible to all connected devices, but only one device can transmit data at a given time. 
Each device consists of a hardware transceiver, which is responsible for data transfer between the bus and the corresponding subsystems.

Control of the 1553 bus is performed by a bus controller (BC) that communicates with a number (up to 31) of remote terminals (RTs) via the 1553 bus. Each RT component contains up to 30 subcomponents.
The BC is the only component assigned the task of initiating information transfer according to a predefined timing and order. 
The BC controls multiple RTs; it polls all of the RTs connected to the 1553 bus. RTs with higher-priority functions (for example, those operating the aircraft control surfaces) are polled more frequently, while RTs with lower-priority functions are polled less frequently. 
To provide control redundancy, a practical system will employ multiple BCs (note that only one device can serve as the BC at a given time). 
There may also be one or more bus monitors (BMs).
A BM is only used to collect data for error analysis; it  is not allowed to take part in data transfers. 

\begin{figure*}[h]
\scriptsize
\centering
\includegraphics[width=0.75\textwidth]{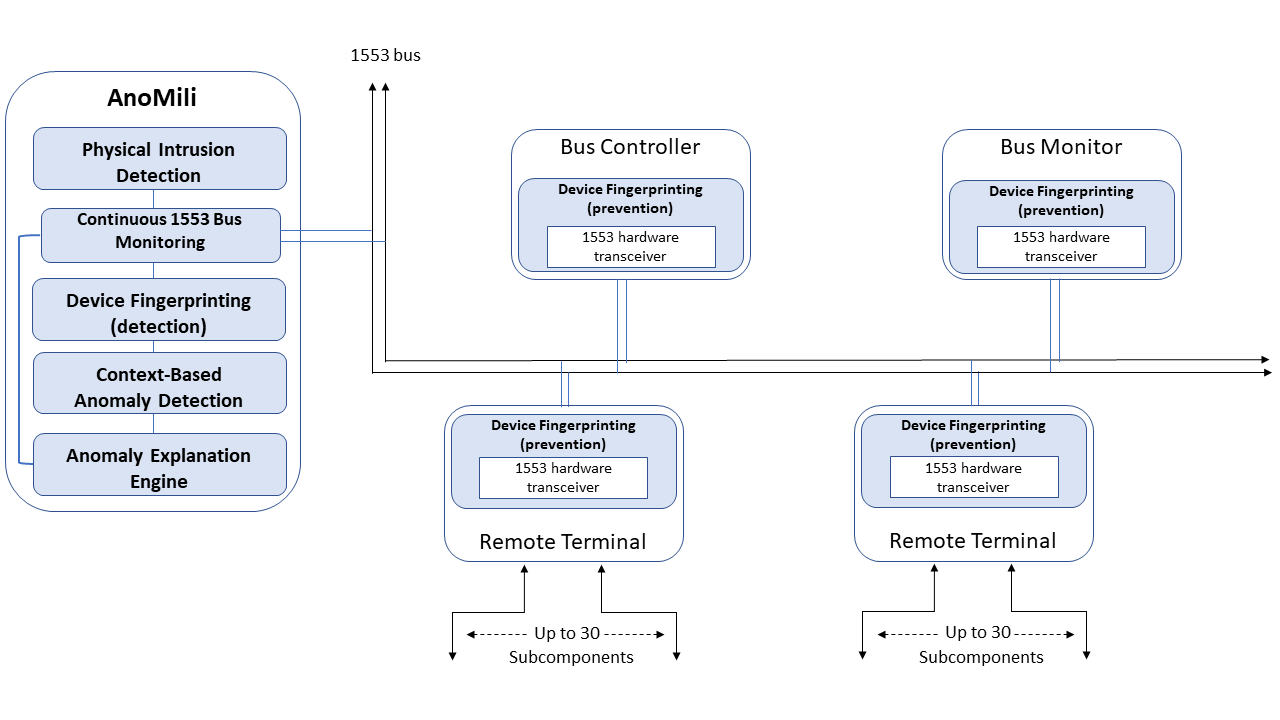}
\caption{General architecture of the MIL-STD-1553 bus integrated with \MethodName (in blue).}
\vspace{0.5cm}
\label{fig:data_bus}
\end{figure*}

Each message transferred on the 1553 bus is organized as a sequence of atomic 20-bit long words. 
As illustrated in Figure~\ref{fig:word_formats}, the standard defines three types of words: command, data, and status. 
Each word begins with a synchronization signal (sync) and ends with a parity bit (p). 
The command word is transmitted by the BC to an RT. 
The command word consists of a terminal address, a transmit/receive bit, a subaddress/mode, and a data word count/mode code. 
The data word contains four bytes of data that is exchanged between two devices. 
The status word is transmitted from an RT back to the BC, immediately after receiving a command, in order to report its state.

The messages transmitted on the 1553 bus are in accordance with the formats in Figure~\ref{fig:information_transfer_formats}. 
There are six message transfer formats: BC to RT, RT to BC, RT to RT, mode command without data word, mode command with data word transmit, and mode command with data word receive. In the BC to RT transfer format, the BC instructs the RT to receive data, while in the RT to BC transfer format, the BC instructs the RT to transmit data. In the RT to RT transfer format, the BC initiates a data exchange between two RTs.

Mode commands are special commands that change the RTs' operation mode. 
Examples of mode commands are timing synchronization requests, self-test requests, and shut down requests. 
As can be seen in Figure~\ref{fig:information_transfer_formats}, the formats of the mode commands are similar to the BC to RT and TR to BC formats, except for: (1) the value of the subaddress/mode field, which is set at 00000b or 11111b; and (2) the value of the word count field, which indicates the operation itself. 
The standard also defines broadcast messages. When they are sent, all RTs suppress their status word transmission to avoid bus collisions. The format of broadcast messages is similar to that of non-broadcast messages, except that the terminal address field is set at 11111b.

Messages in MIL-STD-1553 can be periodic or aperiodic. A major frame is a time frame during which all periodic messages are transmitted at least once (typically 40 to 640 milliseconds long). 
In contrast to the periodic messages, aperiodic messages can be transmitted only once, at a fixed time in the major frame. Since aperiodic messages are event-driven, they are not necessarily transmitted at fixed time intervals.
The time cycles and ordering of the periodic messages, as well as the configuration related to the aperiodic messages, are predefined by the avionic system's designer. 

\section{Threat Model}

We consider an adversary that performs attacks on the 1553 system by injecting malicious messages into the bus using any timing or order. 
In particular, we assume an adversary that: (1) has BC capabilities; (2) is able to sniff the current transmission, in order to learn legitimate patterns; and (3) can associate patterns with their origins and inject spoofed messages accordingly.
Using these capabilities, the adversary can violate the targeted system's:
\begin{itemize}
    \item Integrity - manipulating the original behavior of one or more devices.
    This can be achieved by injecting malicious messages (following a specific timing or order) that contain invalid or incorrect data.
    \item Confidentiality - leaking critical information outside the avionic network. 
    This can be achieved by utilizing compromised devices to establish covert channels or by physically connecting sniffing devices to the network.
    \item Availability - preventing one or more devices from performing their operation or receiving/sending critical data. 
    This can be achieved by manipulating messages to control data routing or cause bus errors.
\end{itemize}

\noindent We present the possible attack surfaces (i.e., attack vectors) in Figure~\ref{fig:attack_surfaces}; malicious messages can be injected into the 1553 bus either by an externally connected device (Figure~\ref{fig:attack_surfaces}, (a)) or via an existing, compromised device (Figure~\ref{fig:attack_surfaces}, (b)). 

\section{Related Work}

The first study that focused on the detection of anomalies in the messages transferred on the 1553 bus was performed by Loiser et al.~\cite{losier2019design}. Their proposed solution uses timing features aggregated in fixed time intervals. The authors profiled benign data transmissions based on manually generated histograms of the values of each timing feature. A time interval is considered anomalous if the average percentage of its difference from a normal baseline exceeds a user-defined anomaly threshold.

An improvement was suggested by Genereux et al.~\cite{genereux2019maidens}. Similar to ~\cite{losier2019design}, the authors only used timing features, but they automated the training process. First, they extracted the features using a sliding time interval, the size of which is optimized automatically according to the inspected traffic; an automated method was used to determine the anomaly threshold.

We observe two significant flaws in the above solutions. First, in both cases, features are extracted for an aggregation of messages rather than for each message individually. This allows an adversary to perform a successful adversarial learning attack. In addition, information loss makes determining the attacker's intent and explaining the detected anomalies infeasible. Second, both solutions are limited to timing features. Therefore, anomalous messages that are transferred (1) at normal timings but out of order, or (2) when the devices transmitting the messages are impersonating their peers (i.e., spoofed messages) cannot be detected.

Stan et al. and Onodueze et al.~\cite{stan2019intrusion,onodueze2020anomaly} presented anomaly detection algorithms that analyze each message individually, utilizing both timing and command features. 
Onodueze et al.~\cite{onodueze2020anomaly} obtained poor results when evaluating different classification methods, since the dataset used for training was highly imbalanced (this is known to cause most classification algorithms to fail or produce poor results); this dataset was collected from a realistic 1553 simulator.
In contrast, Stan et al.~\cite{stan2019intrusion}, who suggested using an unsupervised method, obtained better results by using Markov chains. For evaluation, they set up a real 1553 hardware-based testbed containing one BC and two RTs. 
From the anomaly explanation perspective, one limitation of Markov chains is the need to represent the input instances in a degenerated manner. Each instance is assigned a unique identifier representing a state in the Markov chain; this limits the possibilities for pointing to the most contributing input features to the anomaly.
Another limitation is that Markov-based models are not scalable; adding new instances is not supported without re-collecting a large amount of data and generating the models from scratch.

Stan et al.~\cite{stan2019intrusion} also suggested a mechanism for detecting spoofed messages, which is based on analyzing the voltage signals transferred on the bus. 
They extracted 10 features and used various clustering algorithms to identify the message's origins. 
The proposed spoofing detection method obtained high accuracy when it was evaluated on a bus with just three connected devices; we found that lower accuracy is obtained when there are four or more devices connected to the bus. 
Another drawback of their approach is its inability to detect scenarios in which a silent malicious device is connected to the bus, since the approach depends on the malicious device's transmissions. 
A sniffing device can leak information outside the bus or wait for strategic opportunities to harm the communication. 
In addition, the authors did not consider the fact that the voltage signals transferred on the bus can change over time due to environmental changes, resulting in the need to design a retraining procedure to cope with "concept drift."

The spoofing issues of other standards and protocols used in transportation systems (e.g., ARINC 429 bus~\cite{ji2007design} and CAN bus~\cite{li2008design}) have been widely addressed in the literature. Both the ARINC 429 bus and the CAN bus are serial communication buses that suffer from spoofing vulnerabilities like the 1553 bus.

Some studies examined methods for authenticating the devices that do not require changes to the operating systems or communication protocol. These studies proposed statistical methods for learning and modeling the device communication. However, studies on the CAN bus have demonstrated that such mechanisms can be evaded~\cite{sagong2018cloaking,murvay2020tidal,cho2016fingerprinting}.

Taking the evasion constraint into consideration, other studies proposed methods for detecting spoofed messages on the ARINC 429 bus and CAN bus that are based on analyzing voltage signals~\cite{kneib2018scission,gilboa2020hardware}. A recent study on the CAN bus found that although the software of a device can be compromised, it is difficult to alter the voltage characteristics in a controlled manner~\cite{bhatia2021evading}. 
However, one significant drawback of voltage-based solutions is their need to frequently transition to a retraining mode due to environmental changes. 
This creates an opening for poisoning attacks.

Poisoning attacks against machine learning models have been researched extensively ~\cite{jagielski2018manipulating,biggio2012poisoning,mei2015using}. Rohit et al.~\cite{bhatia2021evading} demonstrated a poisoning attack against voltage-based CAN bus defenses that utilizes a connected malicious device specially designed for this task. We consider a malicious device connection an insider threat, and \MethodName was designed to serve as a defense against this threat; its physical intrusion detection mechanism immediately issues an alert about unauthorized devices maliciously connected to the bus detected when the aircraft starts.

While other methods proposed to cope with spoofing scenarios on the 1553 bus focused only on detection, \MethodName's device fingerprinting mechanism supports both detection and prevention approaches; the detection approach is based on voltage signal analysis and was designed with a retraining procedure, and the prevention approach is based on a wrapper for the 1553 hardware transceiver which actively enforces authorized bus writing in an efficient manner. 
In addition, \MethodName's context-based anomaly detection mechanism demonstrates high performance in detecting anomalous messages; another advantage is its scalability.
Moreover, while all existing solutions for securing the 1553 bus include just a simple anomaly alert mechanism, \MethodName goes beyond this and provides additional information in order to help the user understand the alerts and take the correct action.

\section{\MethodName's Proposed Protection Mechanism}

\begin{figure}[h]
\scriptsize
\centering
\includegraphics[height=0.26\textwidth]{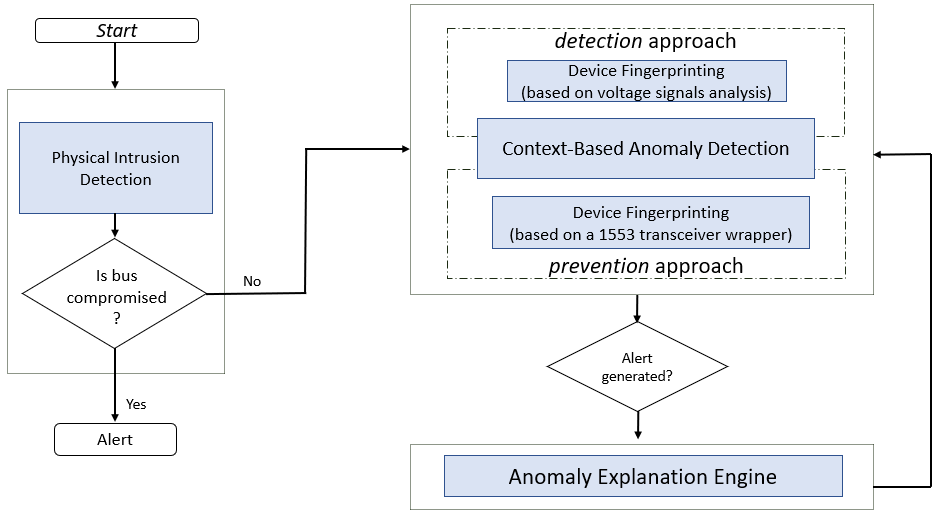}
\vspace{0.2cm}
\caption{\MethodName's high-level architecture. The protection mechanisms are: (1) Physical Intrusion Detection, (2) Device Fingerprinting, and (3) Context-Based Anomaly Detection.
}
\label{fig:high_level}
\end{figure}

\noindent The general architecture of the 1553 bus integrated with \MethodName is demonstrated in Figure~\ref{fig:data_bus}. 
\MethodName is based on continuous monitoring of the messages transferred on the bus and consists of three protection mechanisms and an anomaly explanation engine. 
The protection mechanisms are: the (1) Physical Intrusion Detection, (2) Device Fingerprinting, and (3) Context-Based Anomaly Detection mechanisms.
\subsection{Physical Intrusion Detection}

As illustrated in Figure~\ref{fig:high_level}, when the aircraft starts, the physical intrusion detection mechanism is executed.
This mechanism analyzes the voltage signals transferred on the bus and detects whether an additional device is connected to the bus, at any available entry point.
If a new device is detected, i.e., the 1553 bus is physically compromised, an alert is immediately generated to inform the operational staff.
Available entry points exist on the 1553 bus for maintenance purposes (e.g., for system logs collection and debugging); upon military operations, in legitimate scenarios, no new device is expected to be connected. 

The physical intrusion detection mechanism uses an autoencoder (AE) model that learns the normal patterns of the legitimate devices' voltage signals on the 1553 bus. 
Each voltage signal is digitally represented by a list of \emph{n} voltage samples $v_{1}, v_{2},... ,v_{n}$ collected at a frequency of \emph{V} MHz from the bus during a message transmission. The voltage samples are scaled in the range of [0, 1].
We rely on the fact that each new device connected to the 1553 bus contributes its own resistance and capacitance, modifying the overall electronic characteristics of the bus, and thus affecting the voltage signals of all existing devices. 
Therefore, this mechanism can detect new connected devices even when they are silent, since they modify the electrical behavior of any signal on the bus regardless of an active transmission. 

The AE used for detection is defined with one hidden, fully connected layer containing $\dfrac{n}{2}$ neurons attached with the leaky ReLU activation.

\textbf{Training phase.} To train the AE, we use a dataset that only contains benign data (i.e., voltage signals transferred on the bus when only legitimate devices are connected). 
During the training phase, we first chronologically separate this dataset into a training set (70\%) and a validation set (30\%). 
Then, using the Adam optimizer~\cite{kingma2014adam} initialized with a learning rate of 0.001, we train the AE until the \emph{mean squared error} (MSE) reaches its minimum on the validation set.

When the AE training is complete, a threshold $thresh_{\alpha}$ is determined to discriminate between benign (i.e., voltage signals transferred on the bus when no additional devices are connected) and malicious signals (i.e., voltage signals transferred on the bus when one or more additional devices are connected). $thresh_{\alpha}$ is calculated as the maximum of the samples' maximum of the MSE on the validation set.

\textbf{Detection phase.} During the intrusion detection phase, given a voltage signal transferred on the bus, the AE is executed, and the reconstruction error of the signal is measured. 
If the reconstruction error exceeds $thresh_{\alpha}$, an alert is generated.

If the bus is \textit{not} physically compromised, in the next phase---the monitoring phase---\MethodName starts to continuously monitor the transferred messages on the 1553 bus in order to detect anomalous messages.

\subsection{Device Fingerprinting (Detection)}

The \textit{detection} device fingerprinting mechanism detects unauthorized data transmissions, i.e. spoofing. 
For each legitimate device $d_i$, a CNN-based classifier $CNN_i$ is trained on the voltage signals associated with the device and continuously gets updated to adapt to environmental changes during aircraft operation.
$CNN_i$ provides a binary classification for each voltage signal indicating whether it is associated with the claimed sender ($d_i$) or not.
The input to $CNN_i$ is a list of \emph{n'} voltage samples $v_{1}^{(i)}, v_{2}^{(i)},... ,v_{n'}^{(i)}$ collected at a frequency of $V'$ MHz from the bus during a message transmission by $d_i$. The voltage samples are scaled in the range of [0,1].

This binary classifier consists of three fully connected layers (each with 32 neurons). 
All layers use the ReLU as an activation function. 
A sigmoid layer with a single unit is attached; this layer is aimed at producing the probability that a given example is associated with $d_i$.

\textbf{Training phase.} To induce a binary classifier $CNN_i$ for authenticating device \emph{i}, each signal in the training set is labeled according to the associated sender ('1' if the sender of the voltage signal is device \emph{i} and '0' otherwise).
To address data imbalance, we train each model using the SVM–synthetic minority oversampling technique (SVM–SMOTE)~\cite{chawla2002smote}. 
This technique is responsible for presenting the same proportion of training examples for the positive and negative label for the training subset.

Fore each binary classifier, during the training phase, we use the RMSProp optimizer~\cite{mukkamala2017variants}, with a learning rate of 0.0001, and \emph{binary cross-entropy} is used as the loss function. 
We first chronologically separate the given dataset into a training set (70\%) and a validation set (30\%). 
Then, we train the binary classifier until the loss function reaches its minimum on the validation set.

\textbf{Authentication phase.} Given a voltage signal associated with a transmitting terminal, we extract its identity from the terminal address field specified in the command word and apply the appropriate binary classifier to the signal.
The output returned from the classifier is the probability that the given signal matches the extracted identity. 
If the model output is less than 0.5, an alert is generated.

\textbf{Continuous adaptation to environmental changes.} 
In this work, we assume that environmental changes occur progressively, and accordingly, we use each authenticated signal to retrain the binary classifiers.
Each classifier is retrained given the most recently stored hyperparameters (i.e., neural network's weights, learning rate, and rate decay). 
A single epoch is performed per each authenticated signal.
The physical intrusion detection mechanism ensures that no malicious device is connected to poison the model during retraining. 

\subsection{Device Fingerprinting (Prevention)}
An alternative mechanism for handling spoofing attacks is the \textit{prevention} device fingerprinting mechanism.
This is implemented as a wrapper for the basic 1553 hardware transceiver; this wrapper actively enforces authorized 1553 bus writing based on a given whitelist. 
The whitelist includes all of the possible source addresses of the avionic computers connected to the transceiver. 
The whitelist can be extracted from the 1553 system's designer's notes, or it can be automatically generated during a simple training process. 
If a spoofing attempt is detected, the transmission is blocked, and an alert message $m_{async}$, which contains information regarding the blocked transmission (the transmitting terminal, the spoofed message, and timestamp), is sent asynchronously on the bus.

\subsection{Context-Based Anomaly Detection}

The context-based anomaly detection mechanism receives sequences of consecutive messages transmitted on the bus and detects anomalous messages based on the context they appear in.
This mechanism is based on an Long Short Term Memory (LSTM) AE which learns the normal patterns and behavior, and ensures that each new message complies with the predefined major frame specification as learned during the training phase; given a sequence of consecutive messages as input, this LSTM AE model outputs an abnormal score. This LSTM AE is defined such that its encoder has two layers, where the first has $x$ neurons and the second has $\dfrac{x}{2}$ neurons. 
For each layer, we use the ReLU activation function.
The decoder has a similar structure, although in reverse.

\textbf{Features extracted.} 
In Table \ref{tab:features_extracted}, we present the features extracted from each message. The features include seven command features and one timing feature. Command features can help detect messages that are sent in the incorrect order. Timing features can help detect messages that are sent at suspicious times. The categorical features are one-hot encoded, and the numerical features are normalized. 

\textbf{Training phase.}
Given a parameter \emph{K}, the LSTM AE is trained to reconstruct \emph{K}-length sequences of messages. 
For training, we use a dataset that contains only benign data (i.e., each instance is a sequence of consecutive benign messages). 
During the training phase, we first chronologically separate this dataset into a training set (70\%) and a validation set (30\%). 
Then, using the Adam optimizer initialized with a learning rate of 0.001, we train the LSTM AE until the \emph{mean squared error} (MSE) reaches its minimum on the validation set.

When the AE training is complete, a threshold $thresh_{\beta}$ is determined to discriminate between benign (i.e., sequences of benign messages) and malicious sequences (i.e., sequences of messages whose last message in the sequence is anomalous) . $thresh_{\beta}$ is calculated as the maximum of the samples' maximum of the MSE on the validation set.

\textbf{Detection phase.}
In the detection phase, a message is examined in order to see if it was manipulated, based on the context it appears in. The anomaly detection process is presented in Algorithm \ref{proposed_det_alg}. The input to the algorithm is a sequence of \emph{K}-1 consecutive benign messages that were recently transferred ($lastBenSeq$) and the inspected message ($msg_{t}$).
First, the LSTM AE model is executed given the input sequence (denoted by $input$) set at $lastBenSeq$ and concatenated with $msg_{t}$; the LSTM AE model produces an output denoted by $output$ (lines 5-6). 
Then, the reconstruction error (i.e., the abnormal score, denoted by $mse$) is computed given $input$ and $output$ (line 7). 
If $mse$ is higher than $thresh_{\beta}$, the returned label of $msg_{t}$ is 'Anomalous' (lines 8-9). Otherwise, the returned label of  $msg_{t}$ is 'Benign' (lines 10-11). 
When an anomalous message is detected, an alert is generated.

\begin{table}[h]
\centering
\caption{Extracted features used by the proposed context-based anomaly detection mechanism.}\label{tab:features_extracted}
\scriptsize
\vspace{0.1cm}
\begin{tabular}{|p{2.6cm}|p{1cm}|p{3.4cm}|}
\hline
  \hfil Feature name  & \hfil Values & \hfil Description
 \\ \hline\hline
 \hfil Source Address    & \hfil 0-32      & The address of the device sending the data. If the device is BC, the address is 32.  \\ \hline
 \hfil Source Subaddress  & \hfil 0-32     & The subaddress from which the data is sent in the source device. If the device is BC, the subaddress is 32.   \\ \hline
 \hfil Destination Address  & \hfil 0-32      & The address of the device receiving the data. If the device is BC, the address is 32.  \\ \hline
 \hfil Destination Subaddress & \hfil 0-32  & The subaddress to which the data is received in the destination device. If the device is BC, the subaddress is 32.  \\ \hline
 \hfil Communication Channel & \hfil 0, 1 & The channel on which the message was sent. The value 0 stands for channel A in the protocol, and the value 1 stands for channel B.  \\ \hline
 \hfil Is Mode Command   & \hfil 0, 1  & Specifies whether the command is a mode command or not.   \\ \hline
 \hfil Data Count    & \hfil 0-32  & The number of data words sent in the message frame. If 'Is Mode Command' equals 1, this field specifies the mode command.  \\ \hline
\hfil Time Difference   & \hfil numeric  & The time difference (in microseconds) between the previous message and the current message.   \\ \hline
\end{tabular}
\end{table}

\begin{algorithm}
\caption{Detect anomalies based on context}
\begin{algorithmic}[1]
\Procedure{DetectAnomaly}{$lastBenSeq,msg_{t}$}
    \State $input \leftarrow lastBenSeq || msg_{t}$
    \State $output \leftarrow \Call{lstm\_ae}{input}$
    \State $mse \leftarrow \Call{compute\_mse}{input, output}$
    \If{$mse > thresh_{\beta}$}
         \State \textbf{return} $'Anomalous'$
    \Else
         \State \textbf{return} $'Benign'$
    \EndIf
\EndProcedure
\end{algorithmic}
\label{proposed_det_alg}
\end{algorithm}

\section{\MethodName's Anomaly Explanation Engine}

\begin{figure*}[t!]
\scriptsize
\centering
\begin{minipage}{17.8 cm}
\includegraphics[width=17.8 cm]{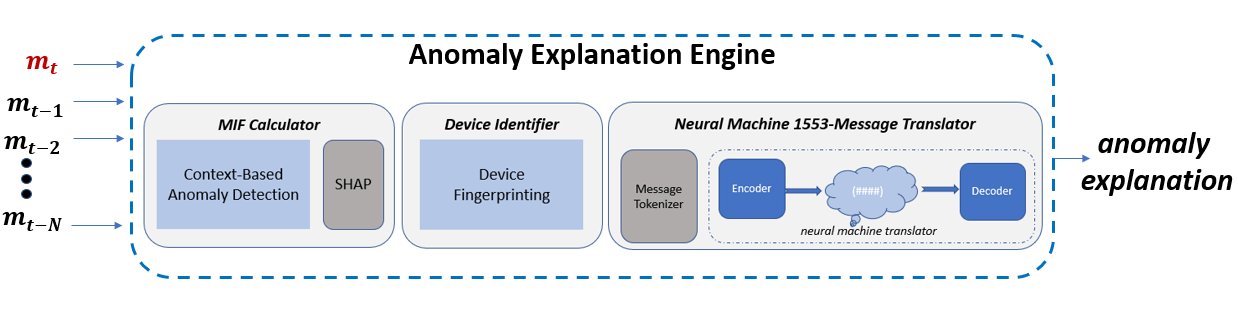}
\centering
\end{minipage}
\vspace{-0.3cm}
\caption{High-level architecture of the proposed anomaly explanation engine.}
\label{fig:anom_exp_fig}
\vspace{0.3cm}
\end{figure*}

The alerts generated by the proposed device fingerprinting and context-based anomaly detection mechanisms trigger the anomaly explanation engine.
The anomaly explanation engine (illustrated in Figure~\ref{fig:anom_exp_fig}) is designed to help \MethodName's users understand the anomalies detected and take the correct action.
Upon detecting an anomalous message $m_{t}, $\MethodName is triggered and receives a sequence of \emph{N} (\emph{N} $\geq$ \emph{K}) consecutive benign messages transferred prior to the anomalous message ($m_{t-N},...,m_{t-2}, m_{t-1}$). 
These messages are used to explain the anomalous message $m_{t}$.
The anomaly explanation engine consists of the following modules: (1) most influential features (MIF) calculator - responsible for identifying the most influential features for an anomaly detected by the context-based anomaly detection mechanism; (2) device identifier - responsible for identifying the attack vector, i.e., which device was compromised and was the sender of the anomalous message $m_{t}$; (3) neural machine 1553-message translator (1553-NMT) - responsible for describing the suspicious event that occurred and what triggered it, by converting (translating) the aircraft operations (as reflected from the 1553 bus) to a human language. A detailed description of each module is provided below.

\subsection{MIF Calculator}
Given an anomalous message $m_{t}$ detected by the context-based anomaly detection mechanism as abnormal, the MIF calculator module identifies the features that contribute the most (i.e., most contributing features) to the abnormal score. 
Despite their high performance on a variety of complex tasks (e.g., anomaly detection~\cite{goldstein2016comparative}), a major drawback of AEs is that their outcomes are hard to explain~\cite{belle2021principles}.

Therefore, we \textbf{locally} approximate the mechanism's outcome by using an interpreted machine learning model (i.e., decision tree) trained in a supervised manner, whose labels are determined based on the abnormal score provided by the context-based anomaly detection mechanism's AE. 

Doing so, creates the opportunity for \MethodName's users to understand the anomalies as follows:
\begin{enumerate} 
    \item Decision tree algorithms provide a straightforward means of explaining predictions~\cite{tsipouras2008automated}; the leaves in decision trees represent class labels, and each input instance is represented by a path from the root to a certain leaf. 
    This path forms a Boolean expression that expresses the relations between input features, making the final decision easy to understand.
    \item The SHAP (SHapley Additive exPlanations) TreeExplainer method~\cite{lundberg2017unified} can be utilized to calculate the most influential features on the model's prediction; each input feature is assigned a score (i.e., a Shapley value) which represents its contribution to the  model's outcome.
    The TreeExplainer method has been proven to be an optimal and efficient method for calculating Shapley values~\cite{lundberg2017unified} for decision tree-based models.
\end{enumerate}

\noindent Given $m_{t}$, the algorithm we use to generate the decision tree $DT_{t}$ as a local approximation is CatBoost~\cite{dorogush2018catboost}. 
CatBoost is an algorithm used for gradient boosting on decision trees, with a straightforward capability of handling categorical features.
$DT_{t}$ is given the \emph{N'} (\emph{K}$\leq$\emph{N'}$\leq$\emph{N}) \emph{K}-length sequences of consecutive benign messages transferred prior to $m_{t}$ and $m_{t}$. 
To avoid an unbalanced training set, we generate additional synthetic examples by applying random valid perturbations to the benign messages. 
We repeat this process until we obtain a balanced training set.

Finally, given an input parameter \emph{F}, the MIF calculator uses the SHAP TreeExplainer method~\cite{lundberg2017unified} to provide the \emph{F} features most contributing  to the abnormal score.

\subsection{Device Identifier}
In the detection approach, this module uses the binary classifiers proposed for device fingerprinting to uniquely identify the real transmitting terminal associated with $m_{t}$. 
This is done using the device fingerprinting mechanism as a building block; each binary classifier is called, given the voltage signal associated with the anomalous message's sender, and the real transmitting terminal is determined based on the maximum score returned by one of the binary classifiers (one for each device connected to the 1553 bus).
In the prevention approach, the real transmitting device is extracted from $m_{async}$ in spoofing attempt scenarios.

\subsection{1553-NMT}
NMT is a state-of-the-art machine translation method~\cite{tiwari2020english}; it is used to translate text from one language to another language. Given a training corpus consisting of a closed group of words/sentences and their translations, an NMT model learns the correlations between words and "understands" short- and long-term meaningful contexts~\cite{wu2016google}. 
Thus, given a new sentence to translate, it is expected to produce a satisfactory translation even when it has not been trained directly. NMTs have been shown to outperform other known machine translation algorithms~\cite{bahdanau2014neural}.

The 1553-NMT module uses a translation model for translating the aircraft operations, as reflected from the 1553 bus, into a human language as the anomalous message $m_{t}$ is transferred; this translation model is generated given the interface control document (ICD) of the 1553 system.
From a practical standpoint, the 1553-NMT module is useful for understanding which aircraft operations occurred immediately \emph{before} the attacker injected $m_{t}$ into the 1553 bus (the attack trigger). 
This is achieved when translating the benign consecutive messages transferred right before $m_{t}$. 
Moreover, given the nature of NMTs, this component is useful for reflecting the attacker's actions as they occur, even when they have not been seen before. This is achieved by translating $m_{t}$ itself.
    
Given a sequence of raw \emph{B'} messages $B'\in\{1, 2,..., B\}$ (\emph{B'}$\leq$\emph{N}), this module translates the sequence into a human language. 
In this work, we utilize the translating framework proposed by Bahdanau et al.~\cite{bahdanau2014neural}. 
This framework is based on a bidirectional LSTM model (referred to as a \emph{translation model}) that consists of an encoder and decoder, which converts source text in one language to destination text in another language. 
We build the 1553-NMT's translation model in two phases: (1) training corpus generation, and (2) translation model training.

\begin{figure}[h]
\scriptsize
\centering
\includegraphics[width=0.5\textwidth]{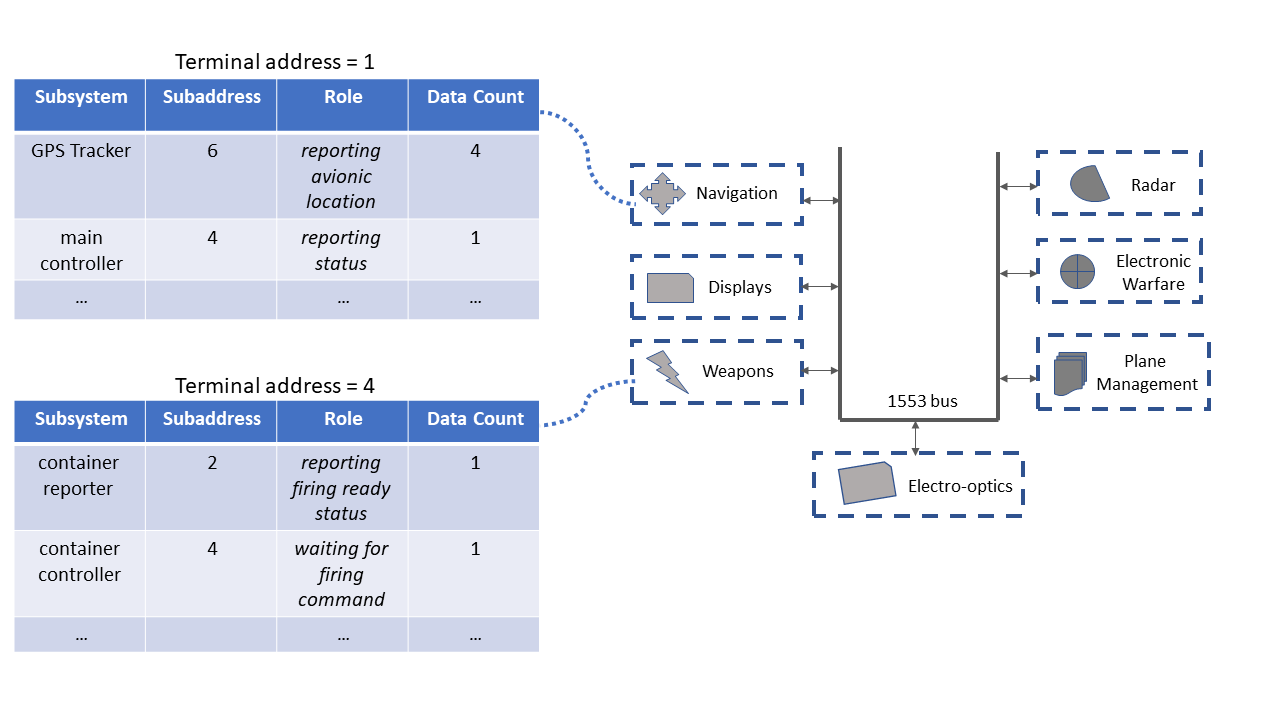}
\caption{Example of 1553 system component subsystem mapping, as extracted from the designer's notes of a 1553 system.}
\label{fig:components_mapping}
\vspace{0.15cm}
\end{figure}

\textbf{Training corpus generation.} 
The training corpus is generated given: (1) the 1553 system's specification, including the ICD, and (2) a set \emph{M} of consecutive messages collected from the 1553 system. The following steps are performed to generate the training corpus:
\begin{enumerate}
    \item Mapping the 1553 system's components: for each 1553 component (associated with a terminal address), we map each of the subcomponents (associated with a terminal subaddress) to its role description in a human language. 
    Specifically, for each subcomponent, we specify whether the subcomponet is waiting for operational commands, reporting internal status, or reporting operational information. 
    For each case, the average word count is specified. 
    An example of such a mapping is illustrated in Figure \ref{fig:components_mapping}. 
    In the example, the navigation component has a subcomponent that reports the location of the aircraft (represented by four data words). 
    The weapon component has a subcomponent that reports 'ready/not ready' firing status (represented by one data word), and another subcomponent that waits for a firing command (represented by one data word). 
    Given the 1553 system's specification and ICD, this mapping table could be generated manually (as we do in this study) or by using neuro-linguistic programming (NLP) techniques~\cite{bandler1982neuro}.
    \item Tokenizing the message features: during tokenizing, each message $m \in M$ represented by an \emph{|m|}-length set of features $f=(f_{1}, f_{2}, ..., f_{|m|})$ is mapped to an \emph{n}-length set of distinct tokens $t=(t_{1}, t_{2}, ..., t_{|m|})$. For a natural number $offset_{i}$, we define $t_{i}$ as $f_{i}+offset_{i}$. To avoid dual meaning, we require that all tokens in $t$ are distinct (for example, we would like to distinguish between a sender and a receiver when we describe a scenario in a human language).
    Let $max_{i}$ be the maximum possible value of the $i$-th feature. For example, the maximum possible value of the source address is 32. To ensure token distinctness, it is sufficient to require that for $i=1$, $offset_{i}=0$ and for each $i, j$ s.t. $i=j-1$, $offset_{i}+max_{i} < offset_{j}$.
    \item Generating the final corpus: during the generation of the final corpus, each message $m \in M$ represented by the \emph{|m|}-length set of tokens $t=(t_{1}, t_{2}, ..., t_{|m|})$ (i.e., source text) is mapped to an \emph{|m|'}-length set of tokens in a human language (i.e., destination text).
    The destination text is determined given the mapping prepared in step 1 above and the mode commands table specified in the 1553 system's ICD.
\end{enumerate}

\noindent For demonstration, in Table \ref{tab:train_corpus}, we present an example of a training corpus generated during the extraction of six message features: $f$=(Source Address, Source Subaddress, Destination Address, Destination Subaddress, Is Mode Command and Data Count). As demonstrated, each sample in the training corpus represents up to \emph{B=3} messages; each message $m_{i}$ is represented by a set of six tokens, where $offset_{1}$=0, $offset_{2}$=33, $offset_{3}$=66, $offset_{4}$=98, $offset_{5}$=131, $offset_{6}$=133.

\textbf{Translation model training.} 
To train the translation model, we use the training method proposed by Bahdanau et al~\cite{bahdanau2014neural}. 
During inference, given a sequence of \emph{B'} messages to translate (\emph{B'} $\leq$ \emph{B}) (each message is represented by an \emph{n}-length set of tokens), we execute the trained translation model to produce a human language description of \emph{B'}.

\begin{table*}[t]
\centering
\caption{Example of a training corpus used to generate the 1553-NMT.} \label{tab:train_corpus}
\begin{tabular}{|l|l|}
\hline
Source Language                             & Destination Language                                                                                                                                            \\ \hline
$m_{1}$ = (1+0, 4+33, 32+66, 32+98, 0+131, 1+133)  & $l_{1}$ = Navigation system reporting status to Bus controller                                                                                       \\ \hline
$m_{2}$ = (2+0, 11+33, 5+66, 7+98, 0+131, 12+133)  & \begin{tabular}[c]{@{}l@{}}$l_{2}$ = Radar system reporting environmental information    to Plane management, steering controller\end{tabular} \\ \hline
$m_{3}$ = (2+0, 7+33, 32+66, 32+98, 0+131, 1+133)  & $l_{3}$ = Radar system reporting status to Bus controller                                                                                            \\ \hline
$m_{4}$ = (32+0, 32+33, 31+66, 0+98, 1+131, 1+133) & $l_{4}$ = Bus controller sending reset command to all RTs                                                                                            \\ \hline
$m_{5}$ = (4+0, 2+33, 32+66, 32+98, 0+131, 1+133)  & $l_{5}$ = Weapons system reporting status to Bus controller                                                                                          \\ \hline
$m_{1}, m_{3}$                                        & $l_{1}$ + ", and then " + $l_{3}$                                                                                                                          \\ \hline
$m_{3}, m_{4}$                                        & $l_{3}$ + ". and then " + $l_{4}$                                                                                                                          \\ \hline
$m_{1}, m_{5}, m_{2}$                                    & $l_{1}$ + ", and then " + $l_{5}$ + ", and then " + $l_{2}$                                                                                                     \\ \hline
$m_{9}, m_{1}, m_{2}$                                    & $l_{9}$ + ", and then " + $l_{1}$ + ", and then " + $l_{2}$                                                                                                     \\ \hline
\end{tabular}
\end{table*}

\section{Experiments and Results}

\begin{figure}[h]
\vspace{-0.35cm}
\hspace*{-0.3cm} 
\includegraphics[height=0.35\textwidth]{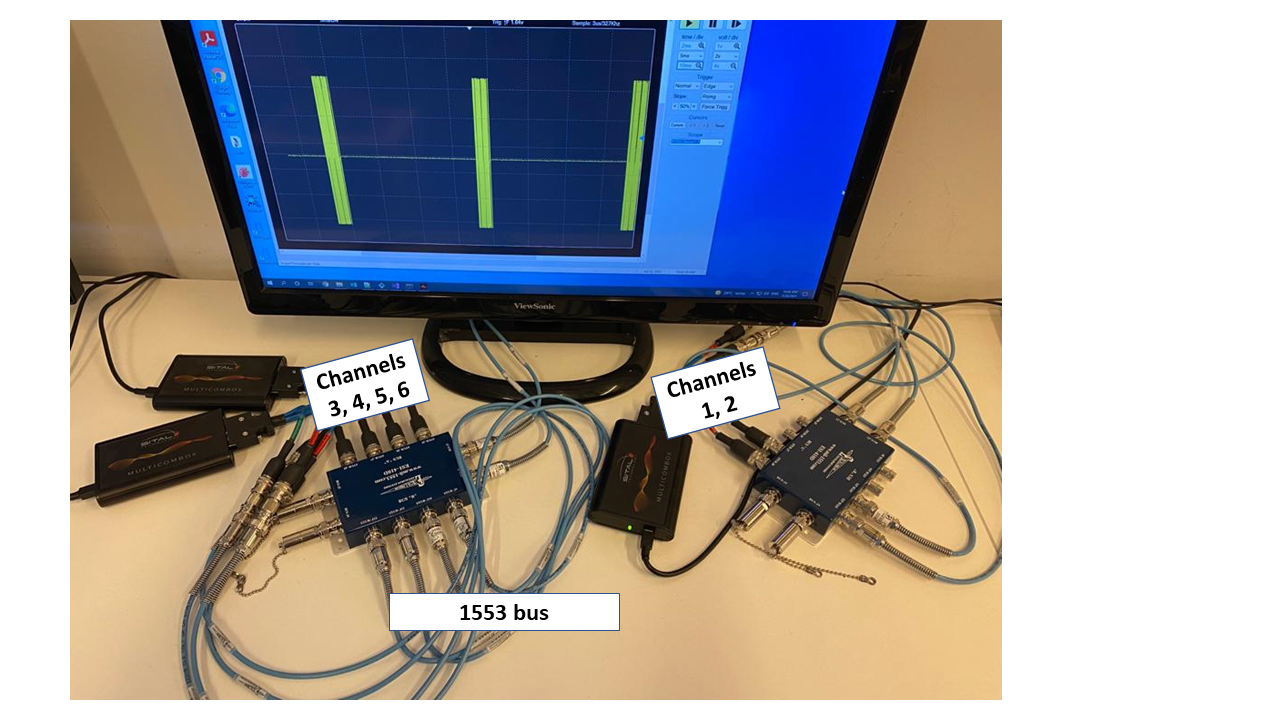}
\caption{Physical components of the testbeds.}
\label{fig:lab}
\end{figure}

\subsection{Experimental Testbed}

To evaluate \MethodName, we set up two 1553 testbeds, each comprised of a real 1553 bus and six physical devices (see Figure~\ref{fig:lab}); to demonstrate the transferability of \MethodName, both of testbeds are set up with an identical 1553 system specification.
Using each testbed, we simulated the components presented in Table~\ref{tab:lab_components}: one BC (on device 1), eight RTs (on devices 2, 3, and 4, by running multi-threaded processes to simulate the RTs), an adversary device with BC capabilities that is unknown to the 1553 system (device 5), and a bus monitor implementing \MethodName (device 6) which includes a built-in 32 MHz 8-bit depth scope and a 1553 message parser.


\begin{table}[]
\centering
\small
\caption{Experimental testbed components.}\label{tab:lab_components}
\vspace{-0.35cm}
\begin{center}
\begin{tabular}{|c|c|c|}
\hline
Component        & Function                 & Device \#          \\ \hline
BC               & Bus controller           & 1                  \\ \hline
RT1              & Communication system     & \multirow{2}{*}{2} \\ \cline{1-2}
RT2              & Plane management system  &                    \\ \hline
RT3              & Weapons system           & \multirow{3}{*}{3} \\ \cline{1-2}
RT4              & Mission system           &                    \\ \cline{1-2}
RT5              & Display system           &                    \\ \hline
RT6              & Flight management system & \multirow{3}{*}{4} \\ \cline{1-2}
RT7              & Navigation system        &                    \\ \cline{1-2}
RT8              & Radar system             &                    \\ \hline
New Device       & Physical intruder        & 5                  \\ \hline
\MethodName & But monitoring           & 6                  \\ \hline
\end{tabular}
\end{center}
\vspace{-0.5cm}
\end{table}

\subsection{Physical Intrusion Detection} 
\textbf{Objective.} Evaluate the mechanism's ability to distinguish between legitimate scenarios (i.e., when no new device is connected to the bus) and physical intrusion scenarios (i.e., when new devices are connected to the bus, regardless of the insertion location or the device's electrical characteristics, considering scenarios in which the new connected device is both passive (sniffing) and active). 

\noindent\textbf{Setup.} For training, we collect hundreds of voltage signals transferred on the bus originating from the legitimate devices (i.e., devices 1-4) when only those devices and device 6 are connected to the bus. 
Each voltage signal is digitally represented as a one-dimensional array that contains \emph{n}=100 numeric values. 
We sample the three sync bits of each word, since these bits are fixed for each word type.

For evaluation, we collect three test sets: (1) 3,000 voltage signals transferred on the bus originating from the legitimate devices when only those devices (i.e., devices 1-4) and device 6 are connected to the bus. (2) 3,000 voltage signals transferred on the bus when the legitimate devices, the adversary device (i.e., device 5), and device 6 are connected to the bus; device 5 is connected alternately to three available points p1, p2, and p3 (1,000 signals are collected per point) and injects messages randomly at bus idle times (to avoid bus collisions, which are easy to detect). (3) This test set is similar to test set 2, however in this case, device 5 is passive.

\noindent\textbf{Results.} Our evaluation results for both testbeds show 100\% accuracy in distinguishing between test set 1 and test set 2.
This shows that the proposed mechanism is able to detect that the bus is physically compromised when an active adversary device is connected (regardless of the connection location).

\begin{figure}[h]
\centering
\vspace{-0.15cm}
\includegraphics[width=0.52\textwidth]{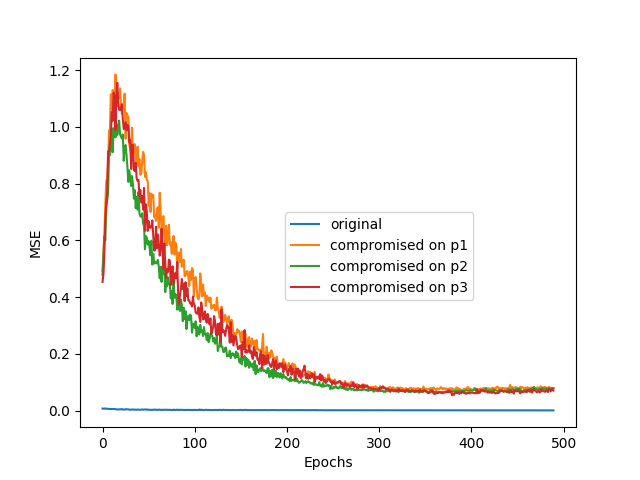}
\caption{The average MSE of the voltage signals originating from the legitimate devices as a function of the number of AE training epochs.}
\label{fig:autoencoder_training}
\end{figure}

To demonstrate the ability of the proposed mechanism to distinguish between test set 1 and test set 3, in Figure~\ref{fig:autoencoder_training} we present the average MSE value obtained (in testbed 1) given the voltage signals originating from the legitimate devices (i.e., test set 1 is referred to as 'original' and test set 3 is referred to as 'compromised on p1') as a function of the number of training epochs. 
As can be seen, there is a statistically significant margin between the reconstruction errors when measuring the voltage signals transferred on the original bus topology and when measuring the voltage signals when the adversary device (i.e., device 5) is connected alternately to one of the three available points on the bus.
A similar phenomenon is observed in testbed 2. 
Our evaluation results for both testbeds show 100\% accuracy in distinguishing between test set 1 and test set 3. 
This shows that the proposed mechanism is able to detect that the bus is physically compromised when a passive adversary device is connected (regardless of the connection location).
All of the above experiments have similar results when the devices swap roles, i.e., when selecting other devices (i.e., devices 1-4) to serve as the adversary device.
These results indicate the robustness of the proposed mechanism to various electrical properties of the devices. Note that to ensure the integrity of the results, the collection of each test set starts after the bus is reset.

\subsection{Device Fingerprinting Evaluation}
\textbf{Objective.} Evaluate the mechanism's ability to detect/prevent spoofing attempts originating from each device connected to the bus, while keeping the amount of incorrect decisions to a minimum.

\noindent\textbf{Setup.} To evaluate the detection mechanism, we collect thousands of voltage signals from each legitimate device (i.e., devices 1-4). 
Each voltage signal is digitally represented as a one-dimensional array that contains \emph{n'}=100 numeric values. 
For training and evaluation, we divide the collected signals chronologically into a training set (50\%), validation set (20\%), and test set (30\%).

Regarding the prevention approach, each device in our setup is equipped with a 1553 hardware transceiver consisting of a transmitter and receiver. 
This transceiver is responsible for receiving/transmitting analog signals from/on the bus; it is responsible for analog to digital conversion and digital to analog conversion. The transmitter is connected to the bus, and the receiver is connected to an FPGA board responsible for encoding/decoding digital data.
The firmware driver of this FPGA board interfaces with a software driver through PCI express, which allows any software module running on the device to perform 1553 communication.
We integrate our spoofing prevention logic into the FPGA driver; our logic only allows authorized bus writing requests originating from software modules running on the device.

\noindent\textbf{Results.} Each CNN-based classifier proposed within the detection approach is evaluated in terms of the false acceptance rate (FAR) and false rejection rate (FRR). 
As can be seen in Table \ref{tab:evaluation_fingerprint}, good results were achieved for all classifiers in both testbeds. 
Regarding the prevention approach, we observe that only legitimate sources could write to the bus. Also, we report that the prevention approach operates with negligible computational overhead (additional details on this are presented in the next subsection). 

\subsection{Context-Based Anomaly Detection} 
\textbf{Objective.} Evaluate (1) \MethodName's ability to detect anomalous messages given a variety of attack scenarios meaningful to typical adversaries, while maintaining a low false alarm rate, and (2) the transferability of the mechanism.

\noindent\textbf{Setup.} In each testbed, we simulate six identical attack scenarios (presented in Table \ref{tab:attack_exp}) in which malicious, harmful messages are injected into the bus. 
For training, in each testbed we simulate normal scenarios and collect thousands of consecutive messages transferred on the bus. 
The training set is used to optimize the $x$ and $K$ parameters to the values of 32 and four respectively. 
For evaluation, we simulate both normal and abnormal scenarios. 
Each test case contains thousands of benign messages and hundreds of malicious messages.
In Table \ref{tab:evaluation_stat}, we present the dataset's statistics for the transferability evaluation.

\noindent\textbf{Results.} Our metrics for evaluation are precision and recall. 
We report perfect results (precision=1 and recall=1) for both testbeds. 
We also report perfect results (precision=1 and recall=1) when training and evaluating on the dataset used by Stan et al.~\cite{stan2019intrusion}. 
In addition, as seen in Table \ref{tab:evaluation_anom}, very good evaluation results were obtained when a model trained in testbed 1 was transferred to testbed 2.

\begin{table}[]
\setlength\tabcolsep{3pt}
\caption{Device fingerprinting mechanism (detection approach) evaluation results for the two testbeds. \label{tab:evaluation_fingerprint}}
\centering
\scriptsize
\begin{center}
\begin{tabular}{|c|c|c|c|c|c|c|c|c|c|}
\hline
\multicolumn{2}{|c|}{Device 1} & \multicolumn{2}{c|}{Device 2} & \multicolumn{2}{c|}{Device 3} & \multicolumn{2}{c|}{Device 4} & \multicolumn{2}{c|}{Device 5}\\ \hline
~FRR~           & ~FAR~        & ~FRR~         & ~FAR~         & ~FRR~         & ~FAR~    & ~FRR~ & ~FAR~ & ~FRR~ & ~FAR~    \\ \hline
\multicolumn{10}{|c|}{Testbed 1}                                                    \\ \hline
 0             & 0          & 0.006           & 0           & 0.024       & 0.004      & 0.002             & 0.005          & 0       & 0.001            \\ \hline
\multicolumn{10}{|c|}{Testbed 2}                                                   \\ \hline
 0             & 0          & 0          & 0           & 0       & 0.005      & 0             & 0         & 0.03       & 0            \\ \hline
\end{tabular}
\end{center}
\end{table}

\begin{table}[]
\centering
\setlength\tabcolsep{1.2pt}
\caption{Context-based anomaly detection mechanism evaluation dataset statistics (when evaluating the mechanism's transferability). \label{tab:evaluation_stat}}
\begin{tabular}{|c|c|c|c|}
\hline
Attack Index & \begin{tabular}[c]{@{}c@{}}Training \\ instances \#\\ (Testbed 1)\end{tabular} & \begin{tabular}[c]{@{}c@{}}Test \\ instances \#\\ (Testbed 2)\end{tabular} & \begin{tabular}[c]{@{}c@{}}\% Malicious \\ instances \\ (Testbed 2)\end{tabular} \\ \hline
1            & \multirow{6}{*}{20,000}                                                        & 3,974                                                                      & 2.81                                                                             \\ \cline{1-1} \cline{3-4} 
2            &                                                                                & 6,529                                                                      & 3.79                                                                             \\ \cline{1-1} \cline{3-4} 
3            &                                                                                & 10,022                                                                     & 3.13                                                                             \\ \cline{1-1} \cline{3-4} 
4            &                                                                                & 5,837                                                                      & 3.77                                                                             \\ \cline{1-1} \cline{3-4} 
5            &                                                                                & 9,877                                                                      & 3.01                                                                             \\ \cline{1-1} \cline{3-4} 
6            &                                                                                & 6,631                                                                      & 3.60                                                                             \\ \hline
\end{tabular}
\end{table}

\begin{table}[]
\centering
\setlength\tabcolsep{6.2pt}
\caption{Context-based anomaly detection mechanism evaluation results in testbed 2 using a model trained in testbed 1. \label{tab:evaluation_anom}}
\begin{tabular}{|c|c|c|}
\hline
Attack Index & Precision & Recall \\ \hline
1            & 1.00       & 1.00    \\ \hline
2            & 0.99      & 1.00    \\ \hline
3            & 0.98      & 1.00    \\ \hline
4            & 0.99      & 1.00    \\ \hline
5            & 1.00      & 1.00    \\ \hline
6            & 0.99      & 1.00    \\ \hline
\end{tabular}
\end{table}

\subsection{Anomaly Explanation Engine}
\textbf{Objective.} Present the explanations generated by the engine with respect to the six simulated attacks (presented in Table \ref{tab:attack_exp}); for each simulated attack we describe the attack vector, the attack trigger, and the attack description.

\noindent\textbf{Setup.} Each explanation is generated given \emph{N=N'}=10, \emph{F}=1, and \emph{B'}=2. More complex explanations result from increasing these arguments.

\noindent\textbf{Results.} In Table \ref{tab:attack_exp}, we present the explanation generated for each simulated attack (note that while the adversary injects the same malicious message a few times, we present the first explanation provided by the engine in testbed 1). Identical explanations are observed for testbed 2. 
For each simulated attack in Table \ref{tab:attack_exp}, we present the output of each anomaly explanation engine module (Figure~\ref{fig:anom_exp_fig}): (1) the malicious message's most influential feature (i.e., the output of the MIF calculator), (2) the real and claimed message origin (i.e., the output of the device identifier), and (3) the attack description (i.e., the output of the 1553-NMT), where the first part of each sentence represents the attack trigger, and the second part represents the attack operation.
As can be observed, each automatically generated explanation does a good job of reflecting the simulated associated attack.

\subsection{Inference Time Measurements}
To demonstrate \MethodName's practicability, in Table \ref{tab:processing_time}, we present the average processing time (in milliseconds) of a single input instance for each of the proposed mechanisms/modules (measured on a 2.11GHz Intel Core i7-8665U processor with 4GB RAM). The time measurement for the 1553-NMT is based on the translation of a single message.
As can be seen, \MethodName protects against malicious activities and explains them within a reasonable amount of time, indicating that \MethodName provides an opportunity for its users to take the right action in response to the anomalies detected.

\begin{table}[]
\centering
\setlength\tabcolsep{7pt}
\caption{Inference time for each \MethodName mechanism/module.}\label{tab:processing_time}
\begin{tabular}{|c|c|}
\hline
Mechanism / Module               & Time (ms) \\ \hline
Physical Intrusion Detection     &        
0.0344        \\ \hline
Device Fingerprinting (detection)           &        
0.0094      \\ \hline
Device Fingerprinting (prevention)           &        
$4.3x10^{-5}$      \\ \hline
Context-Based Anomaly Detection &        
1.2695      \\ \hline
MIF Calculator       &       
0.9883 \\ \hline
Device Identifier (detection)       &       
0.0282   \\ \hline
1553-NMT       &       
10.244   \\ \hline
\end{tabular}
\end{table}

\begin{table*}[t]
\small\addtolength{\tabcolsep}{-0.5pt}
\centering
\scriptsize
\caption{Simulated attack descriptions and their automatically generated explanations on testbed 1.} \label{tab:attack_exp}
\begin{tabular}{|c|c|c|c|cccc|}
\hline
                                                                                  &                                                                              &                                                                                      &                                                                                                                                                                     & \multicolumn{4}{c|}{\cellcolor[HTML]{ACCDEF}Automatically Generated Explanation by AnoMili}                                                                                                                                                                                                                                                                                                                                                                                                                                                \\ \cline{5-8} 
                                                                                  &                                                                              &                                                                                      &                                                                                                                                                                     & \multicolumn{1}{c|}{}                                                                                                          & \multicolumn{2}{c|}{Device Identified}                                                                                                                       &                                                                                                                                                                                                                                             \\ \cline{6-7}
\multirow{-3}{*}{\begin{tabular}[c]{@{}c@{}}Attack\\ Index\end{tabular}} & \multirow{-3}{*}{Attack Vector}                                     & \multirow{-3}{*}{Attack Trigger}                                           & \multirow{-3}{*}{\begin{tabular}[c]{@{}c@{}}Attack Operation: \\ The attacker...\end{tabular}}                                                                  & \multicolumn{1}{c|}{\multirow{-2}{*}{\begin{tabular}[c]{@{}c@{}}Most Influencing \\ Feature on Anomaly\end{tabular}}} & \multicolumn{1}{c|}{Claimed}                                             & \multicolumn{1}{c|}{Real}                                                & \multirow{-2}{*}{\begin{tabular}[c]{@{}c@{}}Attack Description: \\ \ Trigger \& Operation\end{tabular}}                                                                                                                                         \\ \hline
1                                                                                 & \begin{tabular}[c]{@{}c@{}}Compromised RT\\ (Navigation system)\end{tabular} & \begin{tabular}[c]{@{}c@{}}Weapons system\\ ready status update\end{tabular}         & \begin{tabular}[c]{@{}c@{}}... sends \textbf{fake location} \\ \textbf{information} to the \\ cockpit display\end{tabular}                                                            & \multicolumn{1}{c|}{\textbf{Data Count}}                                                                                                & \multicolumn{1}{c|}{\begin{tabular}[c]{@{}c@{}}Navigation \\ system\end{tabular}} & \multicolumn{1}{c|}{\begin{tabular}[c]{@{}c@{}}Navigation \\ system\end{tabular}} & \textit{\begin{tabular}[c]{@{}c@{}}Weapons system reporting \\ firing ready status \\ to Bus controller, \\ and then Navigation system\\ reporting aircraft location \\ to Display system, \\ cockpit display\end{tabular}}                 \\ \hline
2                                                                                 & Compromised BC                                                               & \begin{tabular}[c]{@{}c@{}}Aircraft location\\ update\end{tabular}                   & \begin{tabular}[c]{@{}c@{}}... sends a \textbf{reset command} \\ to the Weapons system\end{tabular}                                                                          & \multicolumn{1}{c|}{\textbf{Mode Command}}                                                                                              & \multicolumn{1}{c|}{BC}                                                           & \multicolumn{1}{c|}{BC}                                                           & \textit{\begin{tabular}[c]{@{}c@{}}Navigation system reporting\\ aircraft location to \\ Bus controller, \\ and then Bus controller \\ sending reset command \\ to Weapons system, \\ container controller\end{tabular}}                    \\ \hline
3                                                                                 & \begin{tabular}[c]{@{}c@{}}Compromised RT\\ (Navigation system)\end{tabular} & \begin{tabular}[c]{@{}c@{}}Detection of user\\ operation\end{tabular}                & \begin{tabular}[c]{@{}c@{}}... \textbf{immediately} replays\\ the steering command \\ sent by the user\end{tabular}                                                          & \multicolumn{1}{c|}{\textbf{Time Difference}}                                                                                           & \multicolumn{1}{c|}{BC}                                                           & \multicolumn{1}{c|}{\begin{tabular}[c]{@{}c@{}}Navigation \\ system\end{tabular}} & \textit{\begin{tabular}[c]{@{}c@{}}Bus controller sending \\ command to \\ Plane management system, \\ steering controller,\\ and then Bus controller\\ sending command to \\ Plane management system,\\ steering controller\end{tabular}}  \\ \hline
4                                                                                 & \begin{tabular}[c]{@{}c@{}}Compromised RT\\ (Radar system)\end{tabular}      & \begin{tabular}[c]{@{}c@{}}Detection of aircraft \\ location update\end{tabular}     & \begin{tabular}[c]{@{}c@{}}... \textbf{floods} the 1553 bus\\ with steering commands\end{tabular}                                                                            & \multicolumn{1}{c|}{\textbf{Time Difference}}                                                                                           & \multicolumn{1}{c|}{BC}                                                           & \multicolumn{1}{c|}{\begin{tabular}[c]{@{}c@{}}Radar\\ system\end{tabular}}       & \textit{\begin{tabular}[c]{@{}c@{}}Navigation system reporting\\ aircraft location \\ to Bus controller,\\ and then Bus controller \\ sending command to \\ Plane management system, \\ steering controller\end{tabular}}                   \\ \hline
5                                                                                 & Compromised BC                                                               & \begin{tabular}[c]{@{}c@{}}Secret data \\ transferred on the\\ 1553 bus\end{tabular} & \begin{tabular}[c]{@{}c@{}}... \textbf{leaks} secret data \\ \textbf{outside} the aircraft\\ by sending commands \\ to the lighting controller\\ (encoding data outside)\end{tabular} & \multicolumn{1}{c|}{\textbf{Destination Address}}                                                                                       & \multicolumn{1}{c|}{BC}                                                           & \multicolumn{1}{c|}{BC}                                                           & \textit{\begin{tabular}[c]{@{}c@{}}Radar system reporting\\ information to \\ Flight management system,\\ main controller,\\ and then Bus controller \\ sending command to \\ Plane management system, \\ lighting controller\end{tabular}} \\ \hline
6                                                                                 & \begin{tabular}[c]{@{}c@{}}Compromised RT\\ (Radar system)\end{tabular}      & No specific trigger                                                                  & \begin{tabular}[c]{@{}c@{}}... transfers \textbf{high amount} \\ \textbf{of data words}, \\ attempting to deny \\ the service of the\\ Flight management \\ system\end{tabular}       & \multicolumn{1}{c|}{\textbf{Data Count}}                                                                                                & \multicolumn{1}{c|}{\begin{tabular}[c]{@{}c@{}}Radar \\ system\end{tabular}}      & \multicolumn{1}{c|}{\begin{tabular}[c]{@{}c@{}}Radar \\ system\end{tabular}}      & \textit{\begin{tabular}[c]{@{}c@{}}Bus controller sending\\ command to \\ Plane management system,\\ steering controller,\\ and then Radar system \\ reporting information to \\ Flight management system, \\ main controller\end{tabular}} \\ \hline
\end{tabular}
\end{table*}


\section{Summary}

In this paper, we propose a novel explainable security system for the 1553 military avionics bus. 
Inspired by the \emph{defense in depth} principle, our system addresses insider threats by detecting devices maliciously connected to the bus. Since we utilize physical side-channels which are independent from malicious data transfer, this can be done immediately when the aircraft starts. 

Next, messages transferred on the bus are continuously monitored. Anomalous messages are detected using the device fingerprinting (both prevention and detection approaches are proposed) and context-based anomaly detection mechanisms. We obtain very good results when evaluating these mechanisms on two real 1553 hardware-based testbeds, as well as when using a dataset consisting of both simulated and real 1553 data that was used in prior work~\cite{stan2019intrusion}.

In order to assist users in understanding the alerts and taking the correct action, we propose an anomaly explanation engine. 
This engine, which is trained given the specifications of the 1553 system, is responsible for identifying the attacker's intent and explaining the detected anomaly in real time. In addition, using the proposed detection mechanisms as building blocks, the anomaly explanation engine can identify the compromised devices and produce an anomaly explanation at a low level of abstraction to be used by technicians or auto-remediation systems.
The experimental results show that the explanations generated by the anomaly explanation engine are consistent with the characteristics of the implemented attacks and the outcomes are very intuitive.

We conclude that our system protects against malicious activities targeting the 1553 military avionics bus and provides good explanations for the anomalies detected within a reasonable amount of time.

Except for the voltage signal-based detection mechanisms, all of the mechanisms proposed in this study are transferable from one 1553 system to another 1553 system without retraining. Regarding the voltage signal-based detection mechanisms, we found that a few minutes of dataset collection and training are sufficient to generate the machine learning models. This indicates the practicability of our system.

\bibliography{main}

\end{document}